# On Hadamard Conjecture

R.N.Mohan[1]
Sir CRR Institute of Mathematics, Eluru-534007, AP, India
Email:mohan420914@yahoo.com

**Abstract:** In this note, while giving an overview of the state of art of the well known Hadamard conjecture, which is more than a century old and now it has been established by using the methods given in the two papers by Mohan et al [6,7].

## 1. Introduction

There are various types of matrices in the literature having distinct properties useful for numerous applications, both practical and theoretical. The famous matrix with orthogonal property is Hadamard Matrix, which was defined by Sylvester in 1867 [11] was studied further by Hadamard in 1893 [4]. The Hadamard matrix H is the given square matrix satisfying $HH^T = nI_n$, which has all the entries in the first row and the first column +1's and the rest of the elements are either +1's or -1's. The inner product of any two rows (columns) is 0. This is called the orthogonal property. Hadamard conjectured that *the Hadamard matrix of order $n \times n$, exists iff $n \equiv 0 \pmod{4}$*. Despite the efforts of several mathematicians, this conjecture remains unproved even though it is widely believed that it is true. This condition is necessary and the sufficiency part is still an open problem. These Hadamard matrices were systematically studied by Paley in 1933 [9]. Then if the inner product is not zero then what are the other possibilities is a question for further exploration. Hadamard matrices have received much attention in the recent past, owing to their wel-known and promising applications. The orthogonal matrices are being studied by Moon Ho Lee, Hou Jia [8], besides giving many applications in communication systems, and Sarukhanyan, Agaian, and Astola in [10] studied the Hadamard matrices and their applications in image analysis, signal processing, fault-tolerant systems, analysis of stock market data, coding theory and cryptology, combinatorial designs and so on.

In this paper the Hadamard conjecture has been discussed in the light of the theory developed in the case of non-orthogonal matrices by Mohan et al [6,7].

The binary matrices are those having only two types of entries, either (0, 1), (1,-1) or literally any two types of entries, but the (+1,-1) binary matrix has been considered in our present discussions. The binary matrices were first studied by Ehlich [2], Ehlich and K. Zeller [3].

------------------------------------

And the nature of non-orthogonal matrix has yet to be studied much by considering (+1,-1) binary matrices. Mohan et al in [6,7] defined three types of non-orthogonal matrices called M-matrices of Type I, II, and III, and have studied their properties. When the matrix is non-orthogonal then it will have orthogonal numbers, which have been formulated in the above three cases. The sum of these orthogonal numbers and some of their properties also have been studied. It is the spirit of these papers that inspired the author to solve this well-known Hadamard conjecture.

Now in the case of Hadamard matrix the orthogonal number is zero. And in the case non-orthogonal matrices Mohan et al [6,7], the orthogonal numbers have been formulated and in one of the cases (M-matrix of Type II), g = n - 4k, k is an integer has been obtained, which paved the way for attempting the Hadamard conjecture that we will study now.

**Definition 1.1.** The **orthogonal number** of a given M-matrix with entries ± 1 is defined as sum of the products of the corresponding numbers in two given rows of the matrix (called inner product of the rows). Consider any two rows $R_l = (r_1, r_2,...,r_n)$ and $R_m = (s_1 s_2,...,s_n)$ and then the orthogonal number denoted by 'g' can be defined as $g = (R_l R_m) = \sum_{i=1}^{n} r_i s_i$ .

In the case of Hadamard matrix as it is an orthogonal matrix then all its orthogonal numbers g's are equal to zero. The Hadamard matrix is defined as a square matrix with entries ±1, such that $HH^T = nI_n$. And the Hadamard conjecture states that *the Hadamard matrix exists iff n ≡ 0 mod (4)*. (Of course for n =2 also Hadamard exists). First we visualize that how this 4 came. Let us consider a matrix M of order n x n, where n is even, with entries +1 or -1 and evaluate the inner product of any two rows.

## 2. An overview

The Hadamard conjecture has been tackled by many an author, for example Tressler [12], worked out one way that n = 4k, using Hamming distance between any two given rows of a binary matrix (0, 1), but it remains at half way to the destination. McCall and Little [5] showed as follows: let $L$ be an integer lattice, and $S$ be a set of lattice points of $L$. We say that $S$ is optimal if it minimizes the number of rectangular sub-lattices of $L$ (including degenerate ones), which contain an even number of points in $S$. We show that the resolution of the Hadamard conjecture is equivalent to the determination of |S| for an optimal set $S$ in a $(4s − 1) × (4s − 1)$ integer lattice $L$. We then specialize to the case of $1 × n$ integer lattices, characterizing and enumerating. Assmus and Key in [1] stated that, "Despite much attention by numerous mathematicians, the central question of existence has not been answered: we do not know whether or not, for every integer *n*, there is an orthogonal 4*n* by 4*n* matrix of plus and minus ones; this, not withstanding that the number of such matrices seems to grow extremely rapidly with *n*, the combinatorial explosion coming perhaps as early as *n* = 7. Still less is known about the classification of Hadamard matrices for general *n* -- but they have been enumerated for *n* < 7. Warwick de Launey and Gordon [14] enunciated the Hadamard conjecture in entirely different terms as, let n be a fixed integer and let **r**(n) denote the largest r for which there is an r×n (1,-1)-matrix H satisfying the matrix equation $HH^T = nI_r$. The Hadamard conjecture states that for n divisible by 4, we have **r** (n) = n. Let ε>0. They have showed that the Extended Riemann Hypothesis and the recent results on the asymptotic existence of Hadamard matrices imply that for n sufficiently large $\mathbf{r}$ (n) > (1/2 –ε) n. Precisely they stated the theorem as follows: Let ε> 0. If the Extended Riemann Hypotheses is true, then for every sufficiently large n ≡ 0 (mod 4) we have $\mathbf{r}$ (n) ≥ $n^{17/22 + \varepsilon}$. The Hadamard conjecture has currently been verified for all n <428. Turyn [13] studied complex Hadamard Matrices and conjectured that the complex Hadamard matrices exist for every order n ≡ 0(mod 2). The truth of this conjecture implies the truth of the Hadamard conjecture also. As the prime aim of our paper, is not expository, but to establish the conjecture in a naive way, much of the literature has not been included.

## 3. Main Result

The Hadamard Matrix H is a square n x n matrix, with entries ±1, and the first row and the first column consist of +1 only, such that $HH^T = nI_n$. It is an orthogonal matrix, where the inner product of any two rows, which is denoted by g is 0.

The Hadamard conjecture states that, "The *Hadamard matrix of order n exists if and only if n≡0 (mod4)." That is n is a multiple of 4. Let n = 4k, where k is an integer. We have to establish that if g = 0, then n = 4k and conversely. Then it is a Hadamard matrix as mentioned above.*

**Proposition 3.1.** Let M be any square matrix of order n x n, with entries +1 or -1 and when n is even, and let the entries in its first row and first column consists of +1's only. Then the orthogonal number between any two rows $R_i$ and $R_j$ where $i \neq j$ is given by 4k-n, where k is the number of unities in the selected set, where $1 \le k \le \dfrac{n}{2}$.

**Proof.** Let $R_i$, and $R_j$ be the two given rows in the M-matrix. We have to calculate their inner product $\langle R_i, R_j \rangle$, where $i \neq j$ and $i \neq 1$. By elementary transformations we can make the row $R_i$ with the first $\dfrac{n}{2}$ elements as +1's and the next $\dfrac{n}{2}$ elements as -1's. The orthogonal numbers of the matrix remain invariant by such elementary transformations. Now consider the other row $R_j$, which has n elements. Divide them in to two sets, the first with $\dfrac{n}{2}$ elements and the second also with $\dfrac{n}{2}$ elements. They can be depicted as follows:

$$R_i = \overbrace{\left(1111...1111...1\right)}^{\frac{n}{2}\,elements} \quad \overbrace{\left(-1-1-1-1...-1-1-1-1\right)}^{\frac{n}{2}\,elements}$$

$$R_j = \left(1\text{-}1\text{-}1..\text{-}1\text{-}1..11\right) \quad \left(11..1\text{-}11\text{-}1\text{-}1\text{-}1\text{-}11..11\right)$$

let +1's be k in number hence -1's are $\frac{n}{2}$-k          +1's are $\frac{n}{2}$-k and -1's are (k) in number

First sets                                    Second sets

Now we evaluate the formula for the orthogonal numbers. Let k be the number of +1's. In the first two sets of the given rows we have (i) the k number of +1's of the row $R_j$ corresponds with k number of +1's in the row $R_i$.(ii) and $\frac{n}{2}$-k number of -1's of the row $R_j$ coincides with same number of +1's of the row $R_i$. Now in the second sets of the two given rows (iii) $\frac{n}{2}$-k number of +1's of the row $R_j$ corresponds with the same number of -1's of the row $R_i$ and (iv) the k number of −1's of the row $R_j$ corresponds with same of -1's in the row $R_i$. Hence we get the formula for the orthogonal number g as follows:

$$g = \left\langle R_i, R_j \right\rangle = k \times 1 \times 1 + ((\frac{n}{2} - k) \times 1 \times (-1)) + ((\frac{n}{2} - k) \times (+1) \times (-1)) + (k)(-1)(-1) = 4k - n \,.$$

When the orthogonal number is 0, then 4k-n = 0. That is n = 4k.

Conversely if n = 4k, if we evaluate the above formula in the same way, when n = 4k then g = 0. Otherwise, when n = 4k, then k = n/4, consequently in the above pattern we obviously see that g = 0. Then will it establish that for n = 4k, there will be Hadamard matrix. Still we have to decide that.

Alternatively, let n ≠ 4k, and then let us see the possibilities or rather absurdities.

I.     Let n = 4k + a, where "a" is any odd number. Since n is even, (n-a) is again odd number and so (n-a)/4 = k, which is absurd as k is an integer and (n-a)/k is not an integer. We can take n = 4k-a also and the same argument holds good.(In fact it is trivial)

II.    Let n = 4k + b, where b is an even number. Then this b should be an even number in two ways, namely either a multiple of 2 or a multiple of 4.

       i).    If b is a multiple of 4, let b=4c, then n=4k+4c=4(k+c), then we got through as n is also a multiple of 4 only, which is required to prove.

              Consider n not as a multiple of 4, but as a multiple of 2. (Since n is a multiple of 4, nothing else is to be proved as it is the required condition).

       ii.    If b is just a multiple of 2, the b = 2x, where x is an odd number say 2z+1. The n = 4k+2x. Where n is also a multiple of 2. Let n = $2n_1$ = 2(2y+1). Then n-2x = 4k, since n is even let n = $2n_1$= 2(2y+1). Then 2(2y+1)-2(2z+1) = 4k. Then (y-z) = k, hence k is an integer.

              Similarly for n = 4k-b, the same argument holds good.

       But this does not help to establish the conjecture. So alternatively we consider formulating for orthogonal numbers again in this situation.

III.   Now consider n just as a multiple of 2, i.e. n = $2n_1$, where $n_1$ is an odd number. If the entities are divided into two halves each is having $n_1$ elements, which is an odd number then

$$R_i = \overset{\text{$n_1$ elements}}{(1111...1111...1)} \qquad \overset{\text{$n_1$ elements}}{(-1-1-1-1...-1-1-1-1)}$$

$$R_j = (1\text{-}1\text{-}1..\text{-}1\text{-}1..11) \qquad (11..1\text{-}11\text{-}1\text{-}1\text{-}1\text{-}11..11)$$

<div align="center"><i>let +1's be k in number hence -1's are $n_1$-k      +1's are $n_1$-k and -1's are (k) in number</i></div>

Now if we evaluate the above formula in this case also since $n_1$ is an odd number let $n_1 = 2\ell + 1$,

Then the orthogonal number g = (k.1.1)+(2$\ell$+1-k)(-1)(+1)+(2$\ell$+1-k)(+1)(-1)+(k.1.1) = 4k-4$\ell$-2. If g = 0 = 4k-4$\ell$-2. Then we have 4(k-$\ell$)-2=0, that is 2(k-$\ell$)=1, Therefore 2k = 2$\ell$+1, which is absurd as an even number can never be equal to an odd number. Hence n should not be a multiple of 2, and hence it should be a multiple of 4.

Thus n ≡ 0 (mod 4) is a needed condition. So the conjecture has not yet been proved completely.

IV. Suppose if we take n = 4k, then will there be Hadamard matrices, if so how to construct them still remains to be established.

To formulate that let us take n = 4k, and consider an n × n matrix with the first row and the first column with +1's. Then let the second row be having the first n/2 = 2k elements as +1's and the second set of n/2=2k elements as -1's. Now we have to fill up the next 4k-2 rows satisfying the condition of orthogonality. Naturally, by haphazard filling any pair of rows may not maintain this orthogonal condition. There will be 4k-1 elements in each row the number of +1's is 2k-1 (leaving +1 of the first column element of that row), and 2k number of -1's. Since the total number of elements are 4k-1, which can be arranged among themselves (4k-1)! ways. But the (2k-1) number of +1's can be arranged among them in (2k-1)! ways and the number of arrangements of (2k) -1's will be (2k)!. But in these two cases as they are same elements the permutations among themselves can not be identified and hence considered as only one permutation in each case. Thus the total number of rows formed with these permutations will be (4k-1)!- (2k-1)!- (2k)! +2 = p (say). Out of these p permutations we require only (4k-2) rows. And from the p permutations each permutation will be a row, and we have to choose (4k-2) rows, i.e.,$pC_{4k-2}$ = $((4k\text{-}1)!\text{-}(2k\text{-}1)!\text{-}(2k)!+2)C_{4k-2}$, which is quite a large number and to fix up the needed combination of rows maintaining the condition of orthogonality is computationally complex and hence can not be decided.

For example let k = 1, we get this number as 5. And among these 5, only three will be different as two rows repeat twice. Thus we remain with 4 rows only which maintain orthogonality.

And as another example for k =2, we have p = 5032 in which we require only 6 rows from among 5032 permutations, i.e., $5032\, C_6$ ways. This shows the impossibility to fix up one pattern with orthogonality, from among the huge number of combinations of rows. In these combinations there will be many pairs of non-orthogonal rows also. And thus even if n = 4k is given we can not formulate the orthogonal matrix. And hence n ≡ 0(mod 4) can not be a sufficient condition. Hence the conjecture established.

**Note 3.1.** For any even value of n, we get the formula for the orthogonal number as 4k-n, and consequently we get distinct orthogonal numbers. But when this g = 0, then orthogonality exists and $HH^T = nI_n$ satisfies and becomes Hadamard Matrix. Conversely when n = 4k, from the above formula g = 0, but in this case of n = 4k, when g =0 zero can not be easily assessed as explained above..

**Note 3.2.** Furthermore as proved that n is not only even it is a multiple of 4. And again when it is a multiple of 4 and if g = 0. This is true whichever pairs of rows that have been taken. And it is not necessary that the elementary transformations be made to have the above pattern. It is for clear perception, this has been taken but now with the same way for any pair of rows, if it has been thought about, then the same formula holds good.

**Note 3.3.** And when the orthogonal numbers are all zero and then $HH^T = nI_n$ is trivial since $\langle R_i, R_i \rangle = n, where\ i = 0, 1, 2, ..., n$ , $\langle R_i, R_j \rangle = 0, where\ i \neq j$ .

**Note.3.4.** The first row and the first column should be all with unities is called a normalization of the matrix. This can be obtained as by replacing any of the rows or columns by their complements, still it remains as a Hadamard matrix. Even if by elementary transformations this pattern can be changed g will remain to be equal to 0.

**Note. 3.5.** The treatment that is being done here is entirely different to that given in Tressler [12], as they have considered the Hamming distance between the two rows, but done halfway only. This is a straight forward, simple and clear proof, facilitating to establish the conjecture.

**Note. 3.6.** We have $\langle R_i, R_i \rangle = n, where\ i = 1, 2, ..., n$ . Because for $\langle R_i, R_i \rangle = n$ , the inner product any row with itself, gives out n only. Also $\langle R_1, R_j \rangle = 0, where\ j = 2, ..., n$

**Note. 3.7.** Now if we consider $HH^T = nI_n$, which is trivial, as the orthogonal number of any two rows is 0 and the orthogonal number of the same rows is n itself.

**Note. 3.8.** In the case when n is odd there is no possibility for g = 0 and hence no orthogonality exists. For further details regarding this type of matrix refer to Mohan et al [6,7], which gave the prime motivation for establishing this conjecture.

**Conclusion:** Thus for all values of n the existence of Hadamard matrices has been evaluated. This ends the episode of Hadamard conjecture, which is more than a century old.

**Acknowledgements:** The author expresses his deep sense of gratitude to Prof. M.G.K.Menon, who is a constant source of inspiration for furthering his research and the authorities of Sir CRR institutions, Secretary G.Subbarao, Principal R.Surya Rao. Also his thanks are due to Prof. Bill Chen, Director Center for Combinatorics, Nankai University, Tianjin-300071, PR China and the Third World Academy of Sciences who facilitated him to further his research studies.

**References**

[1] Assmus, E.F., Key, J.D. (1992). Hadamard matrices and their designs: a coding-theoretic approach. Trans. Amer. Math. Soc, **330**, 269-293.

[2] Ehlich, H. (1964). Determinantenabschätzungen für binäre Matrizen. Math. Z. **83,** 123-132.

[3] Ehlich, H., and Zeller, K. (1962). Binäre Matrizen. Z. Angew. Math. Mech. **42,** T 20-21.

[4] Hadamard, J. (1893). Resolution d'une question relative aux determinants. Bull. des Sciences Mathematiques, **17**, 240--246.

[5] McCall,J., and Little. C. H. C. (1987). The Hadamard Conjecture and Integer Lattices. J. Austral. Math. Soc. Ser. A, **43**, 257-267.

[6] Mohan, R.N., Sanpei Kageyama, Moon Ho Lee, and Gao Yang. (2006). Certain new M-matrices and applications. Submitted to Linear Algebra and Applications on April 10,06 and reference no is LAA # 0604-250B, at this paper can be viewed at http://arxiv.org/abs/cs.DM/0604035, as an e-print.

[7] Mohan, R.N., Moon Ho Lee and Ram Paudal. (2006) An M-matrix of Type III and its Applications. Submitted to Linear Algebra and Applications, on April 11,2006 with refe.No. LAA # 0604-253B and can be viewed at http://arxiv.org/abs/cs.DM/0604044, as an e-print.

[8] Moon Ho Lee, and Hou Jia. (2006). Jacket, Hadamard and allied Orthogonal Matrices for Signal Processing. Lecture Notes Series in Mathematics 2006, Springer Verlag, Germany. (To appear).

[9] Paley, R.A.E.C. (1933). On orthogonal matrices. J.Math.Phys., **12**, 311-320.

[10] Sarukhanyan, H., Agaian, S., and Astola, K.E.J. (2004). Hadamard Transforms. TICSP Series # **23**, ISBN 952-15-1172-9.

[11] Sylvester, J.J. (1867). Thoughts on Orthogonal Matrices, Simultaneous Sign Successions, and Tessellated Pavements in Two or More Colors, with Applications to Newton's Rule, Ornamental Tile-Work, and the Theory of Numbers. *Phil. Mag.*, **34**,461- 475.

[12] Tressler, E. (2004). Survey of the Hadamard Conjecture. Master thesis submitted to Virginia Polytechnic Institute and State University, Blacksburg, VA-24061, USA.

[13] Turyn, R.J. (1970). *Complex Hadamard Matrices, Structures and their Applications.* Gordon and Breach, New York.

[14] Warwick de Launey and Daniel M. Gordon. (2001). A Comment on the Hadamard Conjecture. J. Comb. Th. Ser. A, **95**,180-184.